\documentclass[prd,onecolumn,notitlepage,showpacs,preprintnumbers,amsmath,amssymb,nofootinbib,APS,11pt,superscriptaddress]{revtex4-1}

\usepackage{dcolumn}
\usepackage{bm}
\usepackage[latin1]{inputenc}
\usepackage[spanish,english]{babel}
\usepackage{amsfonts}
\usepackage{amssymb}
\usepackage{graphicx}
\usepackage{dsfont}
\usepackage{hyperref}
\usepackage{color}

\newcommand{\be}{\begin{equation}}
\newcommand{\ee}{\end{equation}}
\newcommand{\bea}{\begin{eqnarray}}
\newcommand{\eea}{\end{eqnarray}}

\begin{document}
\title{{\bf Equivalence of  adiabatic  and DeWitt-Schwinger  renormalization schemes}}

\author{ Adri\'an del R\'{i}o}\email{adrian.rio@uv.es}
\author{Jos\'e Navarro-Salas}\email{jnavarro@ific.uv.es}
\affiliation{Departamento de Fisica Teorica, IFIC. Centro Mixto Universidad de Valencia-CSIC. \\Facultad de Fisica, Universidad de Valencia, Burjassot-46100, Valencia, Spain.}

\date{May 29, 2015}

\begin{abstract}
We prove that adiabatic regularization and DeWitt-Schwinger point-splitting  provide the same result when renormalizing expectation values of the stress-energy tensor for  spin-$1/2$ fields. 
This generalizes the equivalence found  for scalar fields, which is here recovered in a different way. We also argue that the coincidence limit of the DeWitt-Schwinger proper time expansion of the two-point function  agrees  exactly with the analogous expansion defined by the adiabatic regularization method at any  order (for both scalar and spin-$1/2$ fields). We also illustrate the power of the adiabatic method to compute higher order DeWitt coefficients in Fridmann-Lemaitre-Robertson-Walker universes.



\end{abstract}


\maketitle

\section{Introduction}\label{Introduction}

The quantization of the gravitational interaction is one of the most important and difficult problems in theoretical physics. 
Quantum field theory in curved spacetime offers a first step to join Einstein's theory of general relativity and quantum field theory in Minkowski space within a self-consistent and successful framework \cite{parker-toms, birrell-davies}. The  discovery of particle creation  in the
expanding universe 
 \cite{parker66, parker69, parker12} has proven to be of fundamental importance. It implies that particles, perturbations and gravitational waves are created out of the vacuum in the very early universe.  This effect explains the generation of  primordial perturbations in the very early universe \cite{Liddle-Lyth00, Dodelson03} and also  constitutes the driving mechanism to account for the quantum radiance of black holes \cite{hawking74}.
  Within this framework the quantum analysis  of the expectation values of the stress-energy tensor  is of major importance. Since these and other quantities of physical interest are nonlinear in the fields and their derivatives at a single point, the corresponding expectation values diverge in the ultraviolet (UV) regime. This requires renormalization procedures to get rid of the UV infinities in a self-consistent way.  Even for free fields, a curved space-time background introduces additional divergences that are absent in Minkowski space. The renormalization program gets then more involved and a number of methods have been developed to regularize and renormalize expectation values of the stress-energy tensor or other quantities of physical relevance. 
  
The adiabatic subtraction method of regularization  is the most efficient method to carry out the renormalization program in homogeneous cosmological spacetimes. It is especially appropriate in studies in which numerical methods have to be used. It was originally conceived  as a way to overcome the UV divergences in the expectation value of the particle  number operator in Parker's pioneer work on gravitational particle creation   \cite{parker66, parker69}.  It was later generalized by Parker and Fulling \cite{parker-fulling74} to consistently deal with the UV divergences of the  stress-energy tensor of scalar fields in Fridmann-Lemaître-Robertson-Walker (FLRW) space-times. The adiabatic method  naturally identifies  the UV subtracting terms in momentum space, since it is based on the adiabatic asymptotic expansion of the modes characterized by the comoving momentum $\vec{k}$.  It involves a mode-by-mode subtraction process,  in such a way that locality and covariance of the overall renormalization procedure are fully respected. The adiabatic method  is also particularly suitable to scrutinize the primordial power spectrum  in inflationary cosmology \cite{inflation-r} (see also \cite{woodard14}). It has also been used  in the low-energy regime of quantum gravity \cite{agullo-ashtekar}, and more recently, in studies on the breaking of the electric-magnetic duality symmetry in curved space-time \cite{agullo-landete-navarro}.
  
 An alternative asymptotic expansion  (for the two-point function) to consistently identify the subtraction terms  in a generic spacetime was suggested by DeWitt \cite{dewitt75}, generalizing the Schwinger proper-time formalism. The DeWitt-Schwinger expansion was  implemented with the point-splitting renormalization technique  in {\cite{christensen76} and it was  nicely rederived from the local momentum-space representation introduced by Bunch and Parker \cite{bunch-parker}. 
Furthermore, by brute force calculation Birrell \cite{Birrell78} (see also the appendix in \cite{Anderson-Parker87}) checked that  point-splitting and adiabatic renormalization  give the same renormalized stress-energy tensor  
when applied to scalar fields in homogeneous universes.

The extension of the adiabatic regularization method to spin-$1/2$ fields has been achieved very recently \cite{valencia, valencia2}. 
The main difficulty in extending the adiabatic scheme to fermion fields is that the proper asymptotic adiabatic expansion of the spin-$1/2$ field modes does not fit the WKB-type expansion, as happens for scalar fields. However, as shown in \cite{valencia, valencia2}, the method has passed a very nontrivial test of consistency. A major goal of this  paper is to prove that adiabatic regularization and DeWitt-Schwinger point-splitting  will give the same result for the renormalized expectation values of the stress-energy tensor of spin-$1/2$ fields. We base our proof on the  well-known fact that two different methods to compute  $\langle T_{\mu\nu}\rangle$ can differ at most by a linear combination of  conserved local curvature tensors. This result assumes that the  renormalization methods obey  locality and covariance \cite{Waldbook}. Since $\langle T_{\mu\nu}\rangle$ has dimensions of  (length)$^{-4}$ the only candidates are $m^4g_{\mu\nu}$, $m^2G_{\mu\nu}$, $^{(1)}H_{\mu\nu}$ and $^{(2)}H_{\mu\nu}$ (the last two terms can be obtained by functionally differentiating the quadratic curvature Lagrangians $R^2$ and $R_{\mu\nu}R^{\mu\nu}$). It can be seen that the stress-energy tensor only needs  subtraction up to fourth order in the derivatives of the metric  \cite{parker-toms, birrell-davies}, so that higher order contributions need not be considered.

 Therefore, the possible  difference between the expectation values $\langle T_{\mu\nu} \rangle^{Ad} $, computed with adiabatic regularization,  and  $\langle T_{\mu\nu} \rangle^{DS} $, computed with the (DeWitt-Schwinger) point-splitting method,  is parametrized by four dimensionless constants $c_i$, $i=1,... 4$. 
\be \label{ambiguity}  \langle T_{\mu \nu} \rangle^{Ad} - \langle T_{\mu\nu} \rangle^{DS}=   c_1 {^{(1)}}H_{\mu\nu} + c_2 {^{(2)}}H_{\mu\nu} + c_3 m^2 G_{\mu \nu}+ c_4 m^4g_{\mu\nu}  \ . \ee
In our case, the constant $c_4$ is necessarily zero since both  prescriptions lead to a vanishing renormalized stress-energy tensor when restricted to Minkowski spacetime. Moreover, in a FLRW space-time the  conserved tensors ${^{(2)}}H_{\mu\nu}$ and  ${^{(1)}}H_{\mu\nu}$ are not independent, so we can assume without loss of generality  that $c_2\equiv 0$. Therefore, we are left with \be \label{ambiguity}\langle T_{\mu \nu} \rangle^{Ad} - \langle T_{\mu \nu} \rangle^{DS} =   c_1 {^{(1)}}H_{\mu\nu} + c_3 m^2 G_{\mu \nu}  \ . \ee
 Moreover, taking traces in the above relation we get
 \be    \langle T \rangle^{Ad} - \langle T \rangle^{DS} =  -6c_1\Box R-c_3m^2R \ . \ee 
In the massless limit, the classical action of the spin-$1/2$ field is conformally invariant. The trace anomaly calculated with the new adiabatic regularization method has been proved to be in exact agreement with that obtained by other renormalization methods, and in particular with the DeWitt-Schwinger point-splitting method. This implies that  $c_1=0$. Obviously, the same arguments and conclusions apply for a scalar field. The equivalence between both methods is therefore reduced to check that the remaining parameter $c_3$ is also zero. This is actually the most subtle point. 

The comparison between $\langle T^{Ad} \rangle$ and  $\langle T^{DS} \rangle$ can be better studied  by taking into account that, for  spin-$1/2$ fields, $\langle T \rangle = m \langle \bar \psi \psi \rangle$. The  equivalence is then reduced to prove that 
\be \label{psipsiequivalence} ^{(4)}\langle \bar \psi \psi \rangle^{Ad} =  \ ^{(4)}\langle \bar \psi \psi \rangle^{DS} \ , \ee
where $^{(4)}\langle \bar \psi \psi \rangle^{Ad, DS}$ stands for the subtraction terms, up to fourth order in the derivatives of the metric, in the adiabatic and DeWitt-Schwinger expansions respectively. As remarked above, the fourth order is the order required to remove, in general, the UV divergences in the  stress-energy tensor.  To prove (\ref{psipsiequivalence}) and achieve our goal we will make use of the  (Bunch-Parker) local momentum-space representation \cite{bunch-parker} of the two-point function. 
A conceptual advantage of our strategy in comparing both renormalization methods  is that it offers a  better way to spell out their equivalence. In fact, we will also show that the equivalence 
found at fourth order  can be extended to higher order, for both scalar and spin-$1/2$ fields. 

The paper is organized as follows. In Sec. II we consider a similar question for scalar fields. As we have already mentioned, the equivalence between both methods has been checked in \cite{Birrell78, Anderson-Parker87}. We present here an alternative and simpler approach for scalar fields that will allow us to prove the equivalence for spin-$1/2$  fields.  This will be done in Sec. III.  In Sec. IV we extend our results to higher order adiabatic terms. We  will argue that the   coincidence limit of the DeWitt-Schwinger proper time expansion of the two-point function  agrees with the analogous expansion defined by the adiabatic regularization method at any  order. 
Finally, we summarize our conclusions in Sec. V.

\section{Scalar fields}

\subsection{Adiabatic regularization}\label{adiabatic KG}

The general wave equation for a scalar field $\phi$ in a curved space-time is  $(\Box + m^2 +\xi R) \phi =0$, 
where $m$ is the mass of the field and $\xi$ is the coupling of the field to the scalar curvature $R$. If the field propagates in a FLRW space-time [for simplicity we shall assume a spatially flat universe with metric  $ds^2=dt^2- a^2(t)d\vec{x}^2$], it can  be naturally expanded  in the form
\be \phi(x)= \int d^3k \left[  A_{\vec{k}}f_{\vec{k}}(\vec{x}, t) +  A^{\dagger}_{\vec{k}}f^*_{\vec{k}}(\vec{x}, t)\right] \ , 
\ee
where the field modes $f_{\vec{k}} $ are \be f_{\vec{k}}(t,\vec x)=\frac{e^{i\vec{k}\cdot \vec{x}  }    }{\sqrt{2(2\pi)^3a^3(t)}}h_k(t)  \ . \ee
These modes are assumed to obey the normalization condition with respect to the conserved Klein-Gordon product.
This condition translates into a Wronskian-type condition for the modes: $h_k^*\dot h_k - \dot h_k^{*} h_k = -2i$, where the dot means derivative with respect to proper time $t$.
Adiabatic renormalization is based on a generalized WKB-type asymptotic expansion of the modes according to the ansatz
\be \label{kgansatz}  h_k(t) \sim \frac{1}{\sqrt{W_k(t)}}e^{-i\int^t W_k(t')dt'}  \ , \ee
which guarantees  the Wronskian condition.
One then expands $W_k$ in an adiabatic series, in which each contribution is determined by the number of time derivatives of the expansion factor $a(t)$
\be \label{adiabatic}  W_k(t) = \omega^{(0)}(t) + \omega^{(2)}(t)+ \omega^{(4 )}(t) + ... \ , \ee
where the leading term $\omega^{(0)}(t)\equiv \omega(t)= \sqrt{k^2/a^2(t) + m^2}$ is the usual physical frequency. Higher order contributions can be univocally obtained by iteration (for details, see Appendix A),
which come from introducing (\ref{kgansatz}) into the equation of motion for the modes. This adiabatic expansion (\ref{adiabatic}) is  basic to identify  and remove the UV divergences of the expectation values of the stress-energy tensor. 


 The adiabatic expansion of the modes can be easily translated  to an expansion of the two-point function $\langle \phi (x) \phi (x') \rangle \equiv G(x, x')$ at coincidence $x=x'$:
\be G_{Ad}(x,x) = \frac{1}{2(2\pi)^3 a^3} \int d^3 \vec k\,  [ \omega^{-1} + (W^{-1})^{(2)} +  (W^{-1})^{(4)} + ... ] \ . \ee
As remarked above the  expansion must be truncated to the minimal adiabatic order necessary to cancel all UV divergences that appear in the formal expression of the vacuum expectation value  that one wishes to compute. The calculation of the renormalized variance $\langle \phi^2 \rangle$ requires only second adiabatic order, given by  
\be 
(W^{-1})^{(2)}= \frac{m^2\dot a^2}{2a^2\omega^5} + \frac{m^2\ddot a}{4a\omega^5} -\frac{5m^4\dot a^2}{8a^2\omega^7}+\frac{3(\frac{1}{6}-\xi)( \dot a^2 + a\ddot a)}{a^2\omega^3} \ . \ee
The renormalization of the vacuum expectation value of the stress-energy tensor  needs up to fourth adiabatic order subtraction. The corresponding  fourth order contribution $(W^{-1})^{(4)}$ has 30 terms  and can be found in \cite{parker-fulling74}.
Therefore, the adiabatic subtraction terms, truncated to fourth adiabatic order, can be rewritten as 
\bea
\label{adiabaticST2} ^{(4)}G_{Ad}(x,x) =\frac{1}{2(2\pi)^3 a^3}  \int d^3\vec k \,  \left[ \frac{1}{\omega} + \frac{(\frac{1}{6} -\xi)R}{2\omega^3} +\frac{m^2\dot a^2}{2a^2\omega^5} + \frac{m^2\ddot a}{4a\omega^5} -\frac{5m^4\dot a^2}{8a^2\omega^7}+(W^{-1})^{(4)}\right] \ , 
\eea
where we have  taken into account that  $R=6[\dot a^2/a^2 + \ddot a/a]$ in FLRW universes.

We note that only the first two terms in (\ref{adiabaticST2}) are divergent. The remaining  terms  
  can be integrated exactly in momenta producing well-defined finite geometric quantities. Taking into account that $\omega=(\vec{k}^2/a^2 +m^2)^{1/2}$, the integration of the second order adiabatic terms is independent of the mass and gives 
\be  \frac{1}{2(2\pi)^3 a^3} \int d^3 \vec k\,  \left[ \frac{m^2\dot a^2}{2a^2\omega^5} + \frac{m^2\ddot a}{4a\omega^5} -\frac{5m^4\dot a^2}{8a^2\omega^7}\right] =  \frac{R}{288 \pi^2}
\ . \ee 
The integration of the fourth order terms turns out to be also a well-defined geometrical   quantity
\be \label{4th-order-adiabatic}      \frac{1}{2(2\pi)^3 a^3} \int d^3 \vec k \, (W^{-1})^{(4)} = \frac{a_2}{16\pi^2 m^2}   \ , \ee where 
\be a_2= \frac{1}{2}\left[\xi-\frac{1}{6}\right]^2R^2-\frac{1}{6}\left[\frac{1}{5}-\xi\right]\Box R -\frac{1}{180}(R_{\mu\nu}R^{\mu\nu}-R_{\mu\nu\gamma\delta}R^{\mu\nu\gamma\delta}) \ , \ee
is just the coincident point limit $a_2(x) \equiv \lim_{x \to x'} a_2(x, x')$ of the second DeWitt coefficient $a_2(x, x')$ \cite{dewitt75}. (We note that, for our conformally flat space-times, we have $  R_{\mu\nu\gamma\delta}R^{\mu\nu\gamma\delta}= 2R_{\mu\nu}R^{\mu\nu}-\frac{1}{3}R^2$). 
 
In summary, the two-point function for a scalar field at coincidence and at fourth adiabatic order is given by 
\be \label{adiabaticresult} ^{(4)}G_{Ad}(x,x)=      \frac{1}{4 \pi^2 a^3}  \int_0^{\infty} d k k^2 \,  \left[ \frac{1}{\omega} + \frac{(\frac{1}{6} -\xi)R}{2\omega^3}\right] + \frac{R}{288 \pi^2}  +\frac{a_2}{16\pi^2 m^2}\ , \ee 
where the formal divergent term can be understood, for future purposes, as the point-splitting limit 
\be \label{adiabaticresult2}\frac{1}{4 \pi^2 a^3}  \int_0^{\infty} d k k^2 \,  \left[ \frac{1}{\omega} + \frac{(\frac{1}{6} -\xi)R}{2\omega^3}\right] 
\equiv  \lim_{|\Delta \vec x|\to 0}  \frac{1}{4 \pi^2 a^3}  \int_0^{\infty} d k k^2 \frac{\sin(k |\Delta \vec x|)}{k |\Delta \vec x |} \,  \left[ \frac{1}{\omega} + \frac{(\frac{1}{6} -\xi)R}{2\omega^3}\right]  \ .  \ee

\subsection{Local momentum-space representation and DeWitt-Schwinger expansion}\label{localmomentum}

An  alternative asymptotic expansion of the two-point function in momentum space was introduced by Bunch and Parker in \cite{bunch-parker}. It was proposed to aim at extending to curved space the standard momentum-space methods of perturbation theory for interacting  fields in Minkowski space.  This way the standard Minkowskian propagator of a scalar free field in momentum space $(-k^2 +m^2)^{-1}$ is replaced by a series expansion. The Fourier transform leading to local-momentum space is crucially performed with respect to  Riemann normal coordinates $y^{\mu}$ around  a given point $x'$, which constitutes the best possible approximation in curved space to the inertial coordinates of Minkowski space.  
 In contrast to  adiabatic regularization, the method is valid for an arbitrary space-time. It does not serve to (adiabatically) expand the mode functions, which are otherwise highly ambiguous in a general background. The method works directly with the two-point functions, which are regarded as the basic buildings blocks of the renormalization process. 

The covariant expansion of the two-point function $G_{DS}(x, x')$, obeying the equation \bea
(\Box_x+m^2+\xi R) G_{DS}(x, x\, ')=-|g(x)|^{-1/2}\delta(x-x\, ') \ ,
\eea
is defined   in the local-momentum space    
\be \label{BPdefinition}G_{DS}(x, x') = \frac{-i|g(x)|^{-1/4}}{(2\pi)^4}\int d^4k \  e^{iky}\bar G(k)\ , \ee
where $ky\equiv k_0y^0- \vec{k}\vec{y}$ (note that $y^{\mu}(x')=0$), by the series

\bea \bar G(k) &=& \frac{1}{-k^2+m^2} +\frac{(\frac{1}{6}-\xi)R}{(-k^2+m^2)^2}+\frac{i(\frac{1}{6}-\xi)}{2}R_{;\alpha}\frac{\partial}{\partial k_{\alpha}}\frac{1}{(-k^2+m^2)^2}+\frac{1}{3}a_{\alpha \beta}\frac{\partial}{\partial k_{\alpha}}\frac{\partial}{\partial k_{\beta}}{(-k^2+m^2)^{-2}}\nonumber \\ &+&\left[	\left(\frac{1}{6}-\xi\right)^2R^2+\frac{2}{3}a^{\alpha}_{\alpha}\right]\frac{1}{(-k^2+m^2)^3} + ... \ , \eea
where
\be a_{\alpha \beta}= \frac{(\xi-\frac{1}{6})}{2}R_{;\alpha \beta} +\frac{1}{120}R_{;\alpha \beta}-\frac{1}{40}\Box R_{\alpha \beta}-\frac{1}{30}R_{\alpha}^{\  \lambda}R_{\lambda \beta}+\frac{1}{60}R_{\kappa \alpha \lambda \beta}R^{\kappa \lambda}+\frac{1}{60}R^{\lambda \mu \kappa}_{\ \ \ \ \alpha}R_{\lambda \mu \kappa \beta} \ . 
\ee

To compare this local-momentum  expansion  with the adiabatic one introduced in  Sec. \ref{adiabatic KG} we have to convert the momentum-space four-dimensional integrals into  three-dimensional integrals. 
After performing the $k^0$ integration in the complex plane, where the poles in $\bar G(k)$ at $k_0={\pm}\sqrt{\vec{k}^2 +m^2}$ have been displaced in the same way as the analogous Green function in Minkowski space-time, one gets tridimensional integrals. Since all Green functions have the same UV divergences we can perform the contour $k^0$ integration using, for instance, the Feynman prescription for displacing the poles. The result is,
up to fourth adiabatic order
\bea \label{integral}^{(4)}G_{DS}(x, x') &=& \frac{|g(x)|^{-1/4}}{2(2\pi)^3}\int d^3k \, e^{-i (\vec{k}\vec{y}-\sqrt{\vec{k}^2 + m^2}y^0)}     \nonumber \\
&\times&\left[ \frac{a_0 }{(\vec{k}^2 + m^2)^{1/2}} + \frac{a_1(x, x')(1-i y^0 \omega)}{2(\vec{k}^2 + m^2)^{3/2}}  +\frac{3a_2(x, x')(1-i y^0 \omega-(y^0)^2\omega^2/3)}{4(\vec{k}^2 + m^2)^{5/2}}  \right] \\
 &=& \frac{|g(x)|^{-1/4}}{(2\pi)^2|\vec{y}|}\int_0^{\infty} dk\, k \sin(k |\vec{y}|) \,e^{i \sqrt{\vec{k}^2 + m^2}y^0}  \nonumber \\
 &\times&\left[ \frac{a_0 }{(\vec{k}^2 + m^2)^{1/2}} + \frac{a_1(x, x')(1-i y^0 \omega)}{2(\vec{k}^2 + m^2)^{3/2}}  +\frac{3a_2(x, x')(1-i y^0 \omega-(y^0)^2\omega^2/3)}{4(\vec{k}^2 + m^2)^{5/2}}  \right]  \ ,
 \eea 
where $a_0(x,x')\equiv 1$ and, to fourth adiabatic order,  
\bea a_1(x, x') &=& \left[\frac{1}{6}-\xi\right]R(x')  + \frac{1}{2}\left[\frac{1}{6}-\xi \right]R_{;\alpha}(x')y^{\alpha}-\frac{1}{3}a_{\alpha \beta}(x')y^{\alpha}y^{\beta}\ , \nonumber \\ 
 a_2(x, x')&=&\frac{1}{2}\left[\frac{1}{6}-\xi\right]^2R^2(x') +\frac{1}{3}a^{\alpha}_{\alpha}(x')
 \ , \eea
which turn out to be the first  DeWitt coefficients.
 The integrals can be worked out analytically and (\ref{integral}) gives the first three terms in the DeWitt-Schwinger expansion of the two-point function \cite{birrell-davies}
\bea \label{DSG4}
^{(4)} G_{DS}(x,x') & = & \frac{|g(x)|^{-1/4}}{4\pi^2}\left[  \frac{m}{\sqrt{-2\sigma}} K_1(m\sqrt{-2\sigma})+\frac{a_1(x, x')}{2}K_0 (m\sqrt{-2\sigma})+\frac{a_2(x, x')}{4 m} \sqrt{-2\sigma}K_1(m\sqrt{-2\sigma}) \right] \label{result} \ , \ \ \ \ \ \ 
\eea
where $\sigma(x , x')$ is half the square of the geodesic distance between $x$ and $x'$, i.e., $\sigma(x , x')=\frac{1}{2}y_{\mu}y^{\mu}=((y^0)^2- \vec{y}^2)/2$, and $K$ are the modified Bessel functions of second kind.

It is also important to note that  the factor $|g(x)|^{-1/4}$ in the above expressions is evaluated in Riemann normal coordinates with origin at $x'$. The biscalar that reduces to $|g(x)|^{-1/4}$ in arbitrary coordinates is $\Delta^{1/2}(x, x')$, where $\Delta(x, x')$ is  the Van Vleck - Morette determinant, defined as 
\be \Delta(x, x')= -|g(x)|^{-1/2}\det [-\partial_{\mu}\partial_{\nu'}\sigma(x, x')]|g(x')|^{-1/2} \ , \ee
These expressions fit identically with the conventional definition of the DeWitt-Schwinger expansion, as first stressed in \cite{bunch-parker}, which is usually written as
 \be \label{DSdefinition} G_{DS}(x, x') \equiv \frac{\Delta^{1/2}(x, x')}{16\pi^2}\int_0^{\infty}\frac{ids}{(is)^2}\exp\left(-im^2s+\frac{\sigma}{2is}\right)F(x, x'; is) \ , \ee
with
\be F(x, x'; is)= a_0 + a_1(x , x')is + a_2(x, x')(is)^2 + ... \ , \ee
where $a_0=1, a_1, a_2, ... $ are the  DeWitt coefficients. 
To sum up, the Bunch-Parker local momentum-space expansion turns out to be the momentum-space version of the DeWitt-Schwinger expansion of the two-point function. \\

\subsection{Comparison between $^{(4)}G_{Ad}(x,x)$ and $^{(4)}G_{DS}(x,x)$}

To compare the expression (\ref{DSG4}) for $G_{DS}(x, x')$ with the result of  adiabatic regularization  we have  to take the coincident limit $x=x'$ and restrict our analysis  to a spatially flat FLRW universe  $ds^2= dt^2 - a^2(t)d\vec{x}^2$. The comparison is not trivial since in the DeWitt-Schwinger formalism the point-splitting is studied in terms of the geodesic distance $\sigma$. As a first approximation, the normal Riemann coordinates in our FLRW space-time are  $\vec y \approx a\Delta \vec x$. To rigorously  compare with the adiabatic expansion we need the higher order relations between the physical coordinates $(t, \vec x)$ and the normal Riemann coordinates $(y^0, \vec y)$.  The following relations (with $H=\dot a/a$) hold
\cite{brewin} 
\bea
y^0 & = & \Delta t+\frac{1}{2} a^2\Delta \vec x^2 H+ \frac{1}{3} a^2 \Delta \vec x^2\Delta t \left(\frac{R}{12}+H^2\right)+\dots	 \label{F2} \ ,\\
y^i & = & a \Delta x^i \left[1+H \Delta t+\frac{1}{6}a^2\Delta \vec x^2 H^2+\frac{\Delta t^2}{3}\left(\frac{R}{6}-H^2\right)+\dots \right] \ . 	 \label{F3}
\eea
Moreover, 
\be -2\sigma = -\Delta t^2+ a^2 \Delta \vec x^2+a^2 \Delta \vec x^2 H \Delta t+\frac{1}{3}a^2 \Delta \vec x^2\Delta t^2 \left(\frac{R}{6}-H^2\right)+ \frac{a^4\Delta \vec x^4}{12} H^2+\dots  \label{F1} \ , \ee
where, in order to compare to our previous result using the adiabatic regularization, we can just take $\Delta t=0$  without loss of generality and retain the point splitting in $\Delta \vec x$. 

A useful identity for our purposes, using (\ref{F1}) at temporal coincidence  $\Delta t=0$, is
\bea
\frac{1}{-2\sigma} & = & \frac{1}{a^2 \Delta \vec x^2}-\frac{H^2}{12}+O(\Delta \vec x^2) \ . \label{I1}
\eea
Note also that  the factor $|g(x)|^{-1/4}$ in (\ref{result}) is evaluated in Riemann normal coordinates with origin at $x'$ so we can expand $|g(x,x')|^{-1/4} = \Delta^{1/2}(x, x')= 1-\frac{1}{12} R_{\mu\nu}y^{\mu}y^{\nu} + ...$ \ . Another useful relation can be derived using this last result with formulas (\ref{F2}) - (\ref{F3}) [note also that $R_{00}=3\frac{\ddot a}{a}; R_{ii}=-a^2\left(\frac{\ddot a}{a}+2H^2 \right) $], 
\bea
|g(x)|^{-1/4} & = & 1-\left[ 2H^2+\frac{\ddot a}{a} \right] \frac{\sigma}{6}+O(\sigma^{3/2})\ . \label{I2}
\eea

Taking into account (\ref{I1}) and (\ref{I2}), the zeroth order contribution to $^{(4)} G_{DS}(x,x)$ can be reexpressed as 
\bea
\lim_{x\to x\, '}    \frac{|g(x)|^{-1/4} m }{(2\pi)^2 \sqrt{-2\sigma}}K_1(m\sqrt{-2\sigma}) & = & \lim_{ x\to   x\, '}   |g(x)|^{-1/4}\left[-\frac{1}{8\pi^2 \sigma} + O(\log(-\sigma))\right] \\ 
& = & \frac{R}{288 \pi^2}+  \lim_{\Delta \vec x\to 0}   \frac{m }{4 \pi^2 a|\Delta \vec x|}K_1(m\, a|\Delta \vec x|) \\
& = & \frac{R}{288 \pi^2}+  \lim_{\Delta \vec x\to 0}   \frac{1}{4 \pi^2 a^3}\int_0^{\infty} dk k^2 \frac{\sin(k |\Delta \vec x|)}{k |\Delta \vec x|} \frac{1}{\omega} \ .
\eea
Furthermore, the second order contribution is
\bea
 \lim_{x\to x\, '} \frac{|g(x)|^{-1/4}}{4\pi^2}\frac{a_1(x, x')}{2}K_0 (m\sqrt{-2\sigma}) &  = &   \lim_{x\to x\, '}|g(x)|^{-1/4} \times O(\log (-\sigma))\\
  & = &   \lim_{\Delta \vec x\to 0} \frac{1}{4\pi^2}\frac{\left(\frac{1}{6}-\xi\right)R}{2}K_0 (m\, a|\Delta \vec x|)\\
  & = &\lim_{\Delta \vec x\to 0} \frac{1}{4 \pi^2 a^3}\int_0^{\infty} dk k^2 \frac{\sin(k |\Delta \vec x|)}{k |\Delta \vec x| } \frac{\left(\frac{1}{6}-\xi\right)R}{2\omega^3} 
\ , \eea
while the fourth adiabatic term is given by
\bea
 \lim_{x\to x\, '} \frac{|g(x)|^{-1/4}}{4\pi^2}\frac{a_2(x, x')}{4 m} \sqrt{-2\sigma}K_1(m\sqrt{-2\sigma}) = \frac{a_2(x)}{16\pi^2 m^2} \ .
\eea
To sum up, we finally get
\be  \label{BPresult}  {^{(4)}}G_{DS}(x,x) =  \lim_{|\Delta \vec x|\to 0}  \frac{1}{4 \pi^2 a^3}  \int_0^{\infty} d k k^2 \frac{\sin(k |\Delta \vec x |)}{k |\Delta \vec x |} \,  \left[ \frac{1}{\omega} + \frac{(\frac{1}{6} -\xi)R}{2\omega^3}\right] + \frac{R}{288 \pi^2}  +\frac{a_2(x)}{16\pi^2 m^2} \ .   \ee 
By direct comparison with (\ref{adiabaticresult}) and (\ref{adiabaticresult2}) we obtain 
\be \label{equivalenceG4}{^{(4)}}G_{Ad}(x,x)= {^{(4)}}G_{DS}(x,x) \ . \ee


\subsection{Equivalence for $\langle T_{\mu\nu} \rangle $ } 

For the sake of simplicity it is now convenient to restrict ourselves to the case $\xi=1/6$. The reason for which we focus on this particular case is because the spin-1/2 case turns out to be completely analogous, so that it is an illustrative example. In this situation the trace of the stress-energy tensor can be expressed as $\langle T \rangle = m^2 \langle \phi^2 \rangle$. The equivalence $\langle T \rangle^{Ad} = \langle T \rangle^{DS}$, and hence $\langle T_{\mu\nu} \rangle^{Ad}=  \langle T_{\mu\nu} \rangle^{DS}$ (i.e., $c_3=0$, according to the definitions and arguments given in  Sec. \ref{Introduction}), comes directly from the equivalence ${^{(4)}}G_{Ad}(x,x)= {^{(4)}}G_{DS}(x,x)$, since
\be  \langle T\rangle^{Ad}  - \langle T \rangle^{DS}=  m^2\left[{^{(4)}}G_{DS}(x,x)-{^{(4)}}G_{Ad}(x,x)\right] =0 \ . \ee
 
For a general $\xi$, one can compute the stress-energy tensor by acting on the symmetric part of $G(x,x')-{^{(4)}}G(x,x')$ with a certain nonlocal operator, $\langle T_{\mu\nu} (x) \rangle= \lim_{x' \to x} D_{\mu\nu}(x, x') [G(x,x')-{^{(4)}}G(x,x')]$ \cite{birrell-davies, dewitt75, christensen76}.  In Sec. V we have shown the equivalence ${^{(4)}}G_{Ad}(x,x')= {^{(4)}}G_{DS}(x,x')$, which  immediately  implies $\langle T_{\mu\nu} \rangle^{Ad}=  \langle T_{\mu\nu} \rangle^{DS}$ for a general $\xi$.

 \section{Spin-$1/2$ fields}
  
 \subsection{Adiabatic regularization}
 
The first step in  the adiabatic regularization is to define an asymptotic expansion of the field modes. The expansion can be regarded as definitions of approximate particle states in an expanding universe in the limit of infinitely slow expansion.  Spin-$1/2$ fields obey the Dirac equation. In a general background it is given by (see, for instance, \cite{parker-toms, birrell-davies})
\be \left(i{\underline \gamma}^{\mu}\nabla_{\mu}- m\right)\psi =0 \ ,  \ee
where ${\underline \gamma}^{\mu}(x) $ are the spacetime dependent $\underline \gamma$-matrices satisfying the condition $\{{\underline\gamma}^{\mu}, {\underline\gamma}^{\nu}\}= 2g^{\mu \nu}$ and  $\nabla_{\mu} = \partial_{\mu} - \Gamma_{\mu}$ is the covariant derivative associated to the  spin connection $\Gamma_{\mu}$. Let us assume a spatially flat FLRW space-time, with line element $ds^2 = dt^2 - a^2(t) d\vec x^2$.
 The $\underline \gamma$-matrices are related with  the constant Dirac $\gamma$-matrices in Minkowski spacetime by the simple relations: 
$ \underline \gamma^0= \gamma^0 \ , \   \underline \gamma^i (t) = \gamma^i/a(t)$. 
The Dirac equation takes the form 
\be \left[ i\gamma^0\partial_0+ \frac{3i}{2}\frac{\dot a }{a}\gamma^0  +\frac{i}{a}\vec{\gamma}\cdot \vec{\nabla} -m\right]\psi=0 \ . \ee
For our purposes it is convenient to work with the Dirac-Pauli representation of the Minkowskian Dirac matrices 
 \be 
\gamma^0 =
\left( {\begin{array}{cc}
 I & 0  \\
 0 & -I  \\
 \end{array} } \right) \ , 
\hspace{1cm} \vec\gamma = \left( {\begin{array}{cc}
 0 & \vec\sigma  \\
 -\vec\sigma & 0  \\
 \end{array} } \right) \ , 
 \ee
 where the components of $\vec{\sigma}$ are the usual Pauli matrices.
For a given comoving momentum $\vec k$, the basic independent (normalized) spinor solutions are
 \bea
u_{\vec{k}\lambda}(x)=\frac{e^{i \vec{k} \cdot \vec{x}}}{\sqrt{(2\pi)^3 a^3}}
\begin{pmatrix}
h_k^I(t) \xi_{\lambda}(\vec{k})  \\
h_k^{II}(t) \frac{\vec{\sigma}\cdot \vec{k}}{k} \xi_{\lambda}(\vec{k})   
\end{pmatrix},  \label{3}
\eea 
\bea
v_{\vec{k}\lambda}(x)=\frac{e^{-i \vec{k} \cdot \vec{x}}}{\sqrt{(2\pi)^3 a^3}}
\begin{pmatrix}
-h_k^{II*}(t) \xi_{-\lambda}(\vec{k})  \\
-h_k^{I*}(t)  \frac{\vec{\sigma}\cdot \vec{k}}{k}\xi_{-\lambda}(\vec{k})   
\end{pmatrix} \ ,   \label{4d}
\eea
where  $k\equiv |\vec k|$ and $\xi_{\lambda}$ are constant and normalized two-component spinor $\xi_{\lambda}^{\dagger} \xi_{\lambda '}=\delta_{\lambda'\lambda}$. They are chosen to be helicity eigenstates $\frac{\vec \sigma \cdot \vec k}{2k}\xi_{\lambda}(\vec k)= (\lambda/2)\xi_{\lambda}(\vec k)$, where $\lambda/2= {\pm}1/2$. 
 In this decomposition, $h_k^I$ and $h_k^{II}$ are two particular time-dependent functions obeying  the following coupled differential equations:
\bea
h^{II}_k=\frac{i a}{k}(\partial_t+i m)h^{I}_k \ , \hspace{1cm} h^{I}_k=\frac{i a}{k}(\partial_t-i m)h^{II}_k   \ . \label{3b}
\eea
The following self-consistent expansion for the field modes was found in \cite{valencia}
\be h^{I}_{{k}}(t) \sim \sqrt{\frac{\omega + m}{2\omega}} e^{-i \int^{t'} \Omega (t') dt'} F(t) \,\,\,\,\,\, \ ,\,\,\,\,\,\, 
h^{II}_{{k}}(t) \sim \sqrt{\frac{\omega - m}{2\omega}} e^{-i \int^{t'} \Omega (t') dt'} G(t)\ , \label{fermion-ansatz} \ee
where $\omega \equiv \omega^0 \equiv \sqrt{(k/a(t))^2 + m^2}$ is the frequency of the mode and the  time-dependent functions $\Omega (t)$, $F(t)$ and $G(t)$ are expanded adiabatically as
\be \Omega (t) = \sum_{n=0}^{\infty} \omega^{(n)}(t)\ , \ \ \ 
F (t) = \sum_{n=0}^{\infty}  F^{(n)} (t)\ , \ \ \
G (t) = \sum_{n=0}^{\infty}  G^{(n)} (t)
\ . \label{adifexp}\ee
$\omega^{(n)}$, $F^{(n)}$ and $G^{(n)}$ are functions of adiabatic order $n$, which means that they contain $n$ derivatives of the scale factor $a(t)$. We impose $F^{(0)} = G^{(0)} \equiv 1$ at  zeroth order  to recover the Minkowskian solutions for $a(t)=1$. We can solve $\omega^{(n)}$, $F^{(n)}$ and $G^{(n)}$ for $n>1$ by direct substitution of the ansatz (\ref{fermion-ansatz}) into (\ref{3b}) and solving the system of equations order by order. We also have to impose, as an additional order by order requirement,  the normalization condition $|h_k^I(t)|^2+|h_k^{II}(t)|^2 = 1$. For details, see \cite{valencia, valencia2}. The adiabatic series obtained in this way contain  ambiguities. The  ambiguities disappear in the adiabatic expansion of physical vacuum expectation values. 
It is very convenient, for the sake of simplicity, to impose at all adiabatic orders the additional condition $Im G^{(n)} (m) =- Im F^{(n)} (m)$. It  implies that $F^{(n)}(-m)= G^{(n)}(m)$ and removes all the ambiguities. Explicit expressions for the series expansion up to  fourth 
adiabatic order are displayed in \cite{valencia, valencia2}. The algorithm to obtain systematically $\omega^{(n)}$, $F^{(n)}$ and $G^{(n)}$ for any $n$th adiabatic order is shown in Appendix B.

In parallel with the scalar field,  the adiabatic expansion of the spin-$1/2$ field modes can be translated to an expansion of the two-point function $\langle \psi_{\alpha} (x) \bar \psi_{\beta} (x') \rangle \equiv S_{\alpha \beta} (x,x')$ at coincidence $x=x'$. Moreover, since we are  mainly interested  in studying the stress-energy tensor we will restrict our analysis to the trace of the two-point function $\langle \bar \psi (x') \psi (x) \rangle = tr S(x,x')$. Evaluating this at coincidence, the adiabatic expansion up to fourth order is \be  \label{4SAd} tr ^{(4)}S_{Ad}(x,x) =  \frac{-2 }{(2\pi)^3 a^3} \int d^3k\, [ |g_k^{I(4)}|^2-|g_k^{II(4)}|^2 ]  \ ,
\ee
where
\bea g^{I(4)}_{{k}}(t) &\equiv& \sqrt{\frac{\omega + m}{2\omega}} \sum_{n=0}^4F^{(n)} (t)\exp \left[-i \int^{t} \sum_{n=0}^4 \omega^{(n)} (t') dt'\right] \ ,\nonumber \\
g^{II(4)}_{{k}}(t) &\equiv& \sqrt{\frac{\omega - m}{2\omega}} \sum_{n=0}^4G^{(n)} (t)\exp \left[-i \int^{t} \sum_{n=0}^4 \omega^{(n)} (t') dt'\right] \ . \eea

Taking into account that the trace of the stress-energy tensor can be expressed as $\langle T(x) \rangle= m \langle \bar \psi (x) \psi (x) \rangle$, it is very convenient for our purposes to rewrite (\ref{4SAd})  in terms of the expansion for the energy density and pressure \cite{valencia2},
\bea  tr ^{(4)}S_{Ad}(x,x) =\frac{1}{(2\pi)^3 a^3 m} \int d^3k\, \sum_{i=0}^{2}[ \rho_k^{(2i)}-3p_k^{(2i)} ] \ ,
\eea
where,
\bea
\rho_k^{(0)} & = & -2\omega \ , \\
\rho_k^{(2)} & = & - \frac{m^4 \dot{a}^2}{4 \omega^5 a^2} + \frac{m^2 \dot{a}^2}{4 \omega^3 a^2} \label{rho2} \ , \\
p_k^{(0)} & = & -\frac{2\omega}{3} + \frac{2m^2}{3\omega} \ , \\
p_k^{(2)} & = & -\frac{m^2 \dot a^2}{12 \omega^3 a^2}-\frac{m^2 \ddot a}{6 \omega^3 a}+ \frac{m^4 \ddot a}{6 \omega^5 a}+\frac{m^4 \dot a^2}{2 \omega^5 a^2}-\frac{5 m^6 \dot a^2}{12 \omega^7 a^2} \ , 
\eea
and the contribution of the fourth adiabatic order is itself finite and gives
\bea
\frac{1}{(2\pi)^3 a^3 m} \int d^3k\, [ \rho^{(4)}-3p^{(4)} ]=\frac{tr A_2}{16\pi^2 m} 		\label{fourthdirac}\ , 
\eea
where $A_2$ turns out to be  one of the  DeWitt coefficients for spin-1/2 fields at coincidence \cite{bunch-parker, parker-toms} (see next subsection)
\be -A_2(x)=   a_2 (\xi = 1/4){\mathds I}+\frac{1}{48}\Sigma_{[\alpha \beta]}\Sigma_{[\gamma \delta]}R^{\alpha\beta\lambda\xi}R^{\gamma\delta}_{ \hspace{0.3 cm}  \lambda\xi}\ . \ee
In this equation $a_2(\xi=1/4)$ is the DeWitt coefficient for a scalar field with curvature coupling $\xi=1/4$, and 
\bea
\Sigma_{[\alpha \beta]}\equiv \frac{1}{4}\left[\underline \gamma_{\alpha}\underline\gamma_{\beta}-\underline\gamma_{\beta}\underline\gamma_{\alpha} \right] \ .
\eea
Taking into account that
\bea
tr\, \{\Sigma_{[\alpha \beta]}\Sigma_{[\gamma \delta]}\}=g_{\alpha \delta}g_{\beta \gamma}-g_{\alpha \gamma}g_{\beta \delta}
\ , \eea
the term (\ref{fourthdirac}) accounts for the trace anomaly in the massless limit
 \bea
 -\frac{tr A_2}{16\pi^2}= \frac{2}{2880 \pi^2 }\left[-\frac{11}{2}\left(R_{\mu\nu}R^{\mu\nu}-\frac{1}{3}R^2\right)+3\Box R  \right] \ . 
\eea

Let us analyze in detail the lower orders. 
The zeroth order  contribution is easy to handle 
\bea
\frac{1}{(2\pi)^3 a^3 m} \int d^3k\, [ \rho^{(0)}-3p^{(0)} ]=\frac{-m}{\pi^2 a^3 } \int_0^{\infty} dk \, k^2\frac{1}{\omega}  \ .
\eea
However, the second adiabatic order is more subtle. Using the stress-energy tensor  conservation [which is equivalent as imposing the condition $\dot \rho_k^{(n)}+3H p_k^{(n)}=0$], and dimensional regularization, one can eventually arrive at the following expression 
\bea
\frac{1}{(2\pi)^3 a^3 m} \int d^3k\, [ \rho^{(2)}-3p^{(2)} ]= \lim_{n \to 4}    \frac{-m R}{24 \pi^2}\left[ \frac{1}{n-4} +\frac{4}{3}-\log 2 \right] \ .  \label{contr2ad}
\eea
Using now the identity \bea
\frac{4m}{4\pi^2 a^3 } \int_0^{\infty} dk\, k^{2} \frac{R}{24 \omega^3}=\lim_{n \to 4}    \frac{-m R}{24 \pi^2}\left[ \frac{1}{n-4} +1-\log 2 \right] 
\ , \eea
 (\ref{contr2ad}) can be finally expressed as
\bea
\frac{1}{(2\pi)^3 a^3 m} \int d^3k\, [ \rho^{(2)}-3p^{(2)} ]=-4\left[\frac{m R}{288\pi^2}- \frac{m}{4\pi^2 a^3 } \int_0^{\infty} dk\, k^{2} \frac{R}{24 \omega^3}\right] \ . 
\eea
Summing up we have 
\bea
 tr ^{(4)}S_{Ad}(x,x)&=&\frac{-m}{\pi^2 a^3} \int_0^{\infty} dk \, k^2 \left[ \frac{1}{\omega} -  \frac{R}{24 \omega^3}\right] -\frac{4 m R}{288\pi^2}+\frac{tr A_2}{16\pi^2 m} \nonumber \\
 &=&-4 m ^{(2)}G_{Ad}(x,x)|_{\xi=1/4}+\frac{tr A_2}{16\pi^2 m}   \ .\eea

  
\subsection{Local momentum-space representation and DeWitt-Schwinger expansion}

 Following \cite{parker-toms, bunch-parker}, one can construct an asymptotic expansion for the two-point function $\langle \psi (x) \bar \psi (x') \rangle \equiv  S(x, x')$ as follows. Introduce the bispinor ${\cal{G}} (x, x')$ as
 \be S(x, x')\equiv (i{\underline \gamma}^{\mu}\nabla_{\mu} + m) {\cal{G}} (x, x') \ . \ee
 This way we have, as desired, 
\bea
\left(i{\underline \gamma}^{\mu}\nabla_{\mu}- m\right)S(x, x')=\left(\Box +m^2+\frac{1}{4}R \right) [-{\cal{G}}(x, x')]=|g(x)|^{-1/2}\delta(x-x\, ') \ , 
\eea
where we used the identity, $\left({\underline \gamma}^{\mu}\nabla_{\mu}\right)^2=\Box+\frac{1}{4}R$ \cite{parker-toms}. 
We can perform a Fourier  expansion in Riemann normal coordinates around $x'$, as in the scalar case,
\be  {\cal{G}}(x, x')=\frac{-i|g(x)|^{-1/4}}{(2\pi)^4}\int d^4k\  e^{iky}\bar {\cal{G}}(k) \ .  \ee 
The local-momentum expansion for spin-1/2 fields is basically that one for spin-0 fields taking $\xi=1/4$, except for additional spinorial contributions. The detailed expansion can be looked up in \cite{bunch-parker, parker-toms}, and up to fourth adiabatic order reads 

\bea 
\bar {\cal{G}}(k) &=&-\left\{ \frac{{\mathds I}}{-k^2+m^2} -\frac{R \, {\mathds I}}{12(-k^2+m^2)^2}-i\left[\frac{ {\mathds I}}{24}R_{;\alpha}+\frac{1}{12}\Sigma_{[\alpha \beta]}R^{\alpha \beta \hspace{0.15 cm}\lambda}_{\hspace{0.3 cm}\mu\hspace{0.15cm};\lambda}   \right]\frac{\partial}{\partial k_{\alpha}}\frac{1}{(-k^2+m^2)^2} \right. \nonumber\\
 & + & \left[\frac{ {\mathds I}}{3}a_{\alpha \beta}(\xi=1/4)-\frac{1}{48}\Sigma_{[\alpha \beta]}(RR^{\alpha\beta}_{\hspace{0.3 cm}\mu\nu}+R^{\alpha\beta\lambda}_{\hspace{0.45 cm}\mu; \lambda\nu}+R^{\alpha\beta\lambda}_{\hspace{0.45 cm}\nu; \lambda\mu}  ) +\frac{1}{96}\Sigma_{[\alpha \beta]}\Sigma_{[\gamma \delta]}(R^{\alpha\beta\lambda}_{\hspace{0.45cm} \mu}R^{\gamma\delta}_{\hspace{0.3cm} \lambda\nu}+R^{\alpha\beta\lambda}_{\hspace{0.45cm} \nu}R^{\gamma\delta}_{\hspace{0.3cm} \lambda\mu})  \right]  \nonumber \\ 
 &\times&\frac{\partial}{\partial k_{\alpha}}\frac{\partial}{\partial k_{\beta}}{(-k^2+m^2)^{-2}}\nonumber \\ 
 &+&\left.  \left[ \left(\frac{R^2}{288}+\frac{1}{3}a^{\alpha}_{\alpha}(\xi=1/4)\right) {\mathds I}+\frac{1}{48}\Sigma_{[\alpha \beta]}\Sigma_{[\gamma \delta]}R^{\alpha\beta\lambda\xi}R^{\gamma\delta}_{ \hspace{0.3 cm}  \lambda\xi}   \right]  \frac{2}{(-k^2+m^2)^3} + ... \right\} \ , 
\eea

The above expression for the spinor matrix $S(x, x')$ provides an asymptotic expansion of the two-point function $\langle \psi(x) \bar \psi (x') \rangle$, which also turns out to be equivalent to the DeWitt-Schwinger expansion \cite{bunch-parker}. 
Since we are mainly interested in $\langle \bar \psi(x)  \psi (x) \rangle$ we take the   trace of $S(x, x')$ in  formulas above. 
  Taking into account that  $tr\left ( \gamma^{\mu_1}\dots \gamma^{\mu_{2k+1}} \right )=0$, and after performing the 
 contour $k^0$ integration, as in the scalar case,   we obtain
\be  tr \ {^{(4)}}S_{DS}(x,x')= -4m\frac{|g(x)|^{-1/4}}{2(2\pi)^3}\int d^3k e^{-i (\vec{k}\vec{y}-\sqrt{\vec{k}^2 + m^2}y^0)}\left [  \frac{1 }{(\vec{k}^2 + m^2)^{1/2}} - \frac{R(1-i y^0\omega)}{24(\vec{k}^2 + m^2)^{3/2}} +\dots \right ]  \ . \ee 
Restricting now the analysis to a spatially flat FLRW spacetime with metric $ds^2=dt^2 -a^2(t)d\vec{x}^2$ and proceeding in parallel to the scalar case we  get, at coincidence $x=x'$,
\be  tr \ {^{(4)}}S_{DS}(x,x) = -4m\, ^{(2)}G_{DS}(x,x)|_{\xi=1/4}+\frac{tr A_2(x)   }{16\pi^2 m}      \ .   \label{finalBP12}
\ee



\subsection{Comparison between $tr  \ {^{(4)}}S_{DS}(x,x)$ and $tr  \ {^{(4)}}S_{Ad}(x,x)$ and equivalence of $\langle T_{\mu\nu} \rangle$}

It is clear from our previous results that we have a complete agreement between  $tr  \ {^{(4)}}S_{DS}(x,x)$ and $tr  \ {^{(4)}}S_{Ad}(x,x)$ :
\be \label{equiSDSAd} tr  \ {^{(4)}}S_{DS}(x,x)= tr  \ {^{(4)}}S_{Ad}(x,x)= -\frac{m}{\pi^2 a^3} \int_0^{\infty} dk \, k^2 \left[ \frac{1}{\omega} -  \frac{R}{24 \omega^3}\right] -\frac{4 m R}{288\pi^2}+\frac{tr A_2(x)}{16\pi^2 m} \ . \ee
As argued in Sec. I, and taking  into account that $\langle T \rangle = m \langle \bar \psi \psi  \rangle$, the equivalence $\langle T \rangle^{Ad} = \langle T \rangle^{DS}$ for spin-$1/2$ fields, and hence $\langle T_{\mu\nu} \rangle^{Ad}=  \langle T_{\mu\nu} \rangle^{DS}$, can be simply derived from 
(\ref{equiSDSAd}).

\section{Extension to higher orders}
 
 The results obtained in previous sections suggest that the equivalence may go beyond the fourth adiabatic order, i.e., the order required to prove the equivalence of the renormalized expectation values of the stress-energy tensor. We have checked by computed assisted methods that our fundamental relations ${^{(4)}}G_{Ad}(x, x)= {^{(4)}}G_{DS}(x, x)$ and $tr  \ {^{(4)}}S_{Ad}(x,x)= tr  \ {^{(4)}}S_{DS}(x,x)$ are also valid at sixth adiabatic order. In the former case we have
 \be \label{eGDSAd6}{^{(6)}}G_{Ad}(x, x)= {^{(6)}}G_{DS}(x, x)=  \frac{1}{4 \pi^2 a^3}  \int_0^{\infty} d k k^2 \,  \left[ \frac{1}{\omega} + \frac{(\frac{1}{6} -\xi)R}{2\omega^3}\right] + \frac{R}{288 \pi^2}  +\frac{a_2}{16\pi^2 m^2} + \frac{a_3}{16\pi^2m^4} \ , \ee 
where the value obtained for the purely sixth adiabatic order contribution  matches exactly with the third order DeWitt coefficient $a_3$. The  general expression for the coefficient $a_3$, which has 28 terms, was first obtained in \cite{sakai, gilkey}, and can also be found in \cite{parker-toms} (see Chapter 3, Sec. 3.6). We note that the  above agreement  is consistent with that found in \cite{Kaya, Matyjasek} in terms of the sixth order adiabatic approximation for the renormalized stress-energy tensor of scalar fields.

We have also tested the equivalence at sixth adiabatic order for spin-$1/2$ fields
\be \label{eSDSAd6}tr  \ {^{(6)}}S_{DS}(x,x)= tr  \ {^{(6)}}S_{Ad}(x,x)= -\frac{m}{\pi^2 a^3} \int_0^{\infty} dk \, k^2 \left[ \frac{1}{\omega} -  \frac{R}{24 \omega^3}\right] -\frac{4 m R}{288\pi^2}+\frac{tr A_2}{16\pi^2 m} + \frac{tr A_3}{16\pi^2 m^3} \ . \ee
The adiabatic method produces the result
\be tr A_3 = -\frac{2}{21 }\frac{\dot  a^4}{a^4}\frac{\ddot a}{a}  +\frac{8}{21}\frac{\dot a ^2}{a^2}\frac{\ddot a ^2}{a^2}-\frac{4 \ddot a^3}{45 a^3}+\frac{2 }{21}\frac{ \dot a^3}{a^3}\frac{\dddot a}{a}-\frac{2 }{5 }\frac{\dot a\ddot a\dddot a}{a^3} -\frac{\dddot a^2}{210 a^2}-\frac{\dot a^2 \ddddot a}{15a^3}+\frac{\ddot a \ddddot a}{105 a^2}	+\frac{2 \dot a  a^{(5)}  }{35 a^2}+\frac{ a^{(6)} }{70 a}	\label{trA3}		\ , \ee
where we have used the obvious notation [$a^{(n)} \equiv \frac{d^{n}}{dt^n} a$]. We have  checked that (\ref{trA3}) agrees with the third order DeWitt coefficient for fermions \cite{avramidi}, 
\bea
-A_3(x) & = & a_3(\xi=1/4) \mathds I-\Sigma^{[a b]}\Sigma^{[c d]}\left[  \frac{R}{576}R_{ab\mu\nu}R_{cd}^{\hspace{0.3 cm}\mu\nu} +\frac{1}{720} R_{ab\mu\nu}^{\hspace{0.7 cm}; \mu}R_{cd\alpha}^{\hspace{0.55cm}\nu;\alpha }+\frac{1}{120}R_{ab\mu\nu}R_{cd\alpha}^{\hspace{0.5cm}\nu;\alpha\mu }  \right. \nonumber\\
 &  &  \left.  +\frac{1}{180}R_{ab\mu\nu;\alpha} R_{cd}^{\hspace{0.5cm}\mu\nu;\alpha } +\frac{1}{72} R^{\alpha\beta}R^{\hspace{0.4cm}\mu}_{ a b \hspace{0.25cm} \alpha} R_{ c d \mu \beta}  -\frac{1}{240}R^{\mu\nu\alpha\beta}R_{ab\mu\nu}R_{cd \alpha\beta} \right] \nonumber\\
 &&+\frac{1}{80} \Sigma^{[a b]}\Sigma^{[c d]}\Sigma^{[e,f]} R_{ab\mu\nu}R^{\hspace{0.4cm}\nu}_{cd\hspace{0.4cm}\gamma}R_{ef}^{\hspace{0.4cm}\gamma\mu} \ .
\eea 
We have also checked that this contribution is  consistent with the purely sixth adiabatic order of the renormalized stress-energy tensor that has been reported in \cite{Matyjasek} (see also \cite{torrenti}).  
 
Taking into account all this, it seems natural to argue that relations (\ref{eGDSAd6}) and (\ref{eSDSAd6}) are also valid for an arbitrary $n$th order,  since both adiabatic and DeWitt-Schwinger methods provide a series expansion in which each contribution is univocally derived from some well-defined recursion relations using the first order terms as seeds for iteration. We have explicitly seen that the leading sixth order contributions agree, so it is very likely that higher order terms will agree as well. The calculation of the fourth and higher order DeWitt coefficients  has been an elusive problem for a long time. The formal solution, given by a very involved recursion mechanism, was given in \cite{avramidi}.  To show the power of the adiabatic method for cosmological space-times,  and also as an illustrative example, we have  easily worked out the explicit form of  the fourth  DeWitt-Schwinger coefficient $a_4(x)$  using (\ref{A}). It is given in Appendix C. 
 
 \section{Extension to separate points}
 
 Finally, we would like to analyze the two-point functions, expanded up to a given adiabatic order, at separate points. The calculations are much more involved. We illustrate here explicitly the equivalence found at fourth adiabatic order for scalar fields.
The adiabatic scheme provides the following result:
\bea
\label{adiabaticST2sp} ^{(4)}G_{Ad}((t, \vec x), (t, \vec x')) &=&\frac{1}{2(2\pi)^3 a^3}  \int d^3\vec k \, e^{i\vec k \Delta \vec x}  \left[ \frac{1}{\omega} + \frac{(\frac{1}{6} -\xi)R}{2\omega^3} +\frac{m^2\dot a^2}{2a^2\omega^5} + \frac{m^2\ddot a}{4a\omega^5} -\frac{5m^4\dot a^2}{8a^2\omega^7}+(W^{-1})^{(4)}\right] \nonumber \\
&=& \frac{m}{4\pi^2 a |\Delta \vec x|}K_1(am|\Delta \vec x|) + \frac{(\frac{1}{6} - \xi)R}{8\pi^2}K_0(am|\Delta \vec x|) \nonumber \\
&+& \frac{R}{288 \pi^2}(ma|\Delta \vec x|)K_1(am|\Delta \vec x|) - \frac{H^2}{96\pi^2}(am|\Delta \vec x|)^2K_0(am|\Delta \vec x|) \nonumber\\ 
& + & \frac{1}{2(2\pi)^3 a^3}  \int d^3\vec k \, e^{i\vec k \Delta \vec x}   (W^{-1})^{(4)}  \ , 
\eea 
where
\bea
\frac{1}{2(2\pi)^3 a^3}  \int d^3\vec k \, e^{i\vec k \Delta \vec x}   (W^{-1})^{(4)}   = \hspace{15 cm}  \\
  \frac{K_{0}(am|\Delta \vec x|)}{\pi^2} \left\{ -\frac{7|\Delta \vec x|^4 a^4 H^4}{5760  } -\frac{11 m^2|\Delta \vec x|^4 a^4 H^2 \ddot a}{5760  a}- \frac{|\Delta \vec x|^2\xi a H^2 \ddot a}{4  a}+ \frac{43|\Delta \vec x|^2 a^2 H^2 \ddot a}{960  a}			\right. \hspace{5 cm}\nonumber\\
\left. +\frac{3 |\Delta \vec x|^2  \ddot a^2}{320  a^2} -\frac{|\Delta \vec x|^2 \xi \ddot a^2}{16 a^2}	+\frac{7 |\Delta \vec x|^2}{960 }H \frac{\dddot a}{a} -\frac{|\Delta \vec x|^2 \xi H \dddot a}{16 a} -\frac{|\Delta \vec x|^2 a^2 \ddddot a}{960 a} \right\} \hspace{5 cm}\nonumber\\
+   \frac{K_{1}(am|\Delta \vec x|)}{\pi^2} \left\{ H^4\left[ \frac{|\Delta \vec x| a}{32 m }- \frac{3 \xi |\Delta \vec x| a}{8 m } + \frac{9\xi^2 |\Delta \vec x| a}{8 m }	- \frac{m |\Delta \vec x|^3 a^3}{180  }+ \frac{m |\Delta \vec x|^3  a^3 \xi}{32  }+ \frac{m^3 |\Delta \vec x|^5 a^5}{4608 }	\right]		\right. \hspace{4 cm} \nonumber\\
\left. H^2 \frac{\ddot a}{a}\left[ \frac{m |\Delta \vec x|^3 a^3}{2880  }+ \frac{m \xi |\Delta \vec x|^3a^3 }{32  } + \frac{29 |\Delta \vec x| a}{240 m }	- \frac{17 |\Delta \vec x| a \xi}{16 m  }+ \frac{9  |\Delta \vec x|  a \xi^2}{4 m  }  \right] \right. \hspace{6 cm} \nonumber\\
 \frac{\ddot a^2}{a^2}\left[ \frac{3 |\Delta \vec x| a}{160 m  }- \frac{-5 \xi |\Delta \vec x| a }{16 m  } + \frac{9 |\Delta \vec x| a \xi^2}{8 m }	+ \frac{m |\Delta \vec x|^3 a^3 \xi}{640   } \right]	+\frac{\dddot a}{a} H\left[ \frac{3  |\Delta \vec x|  a }{80 m  } + \frac{3  |\Delta \vec x|  a \xi }{16 m  }+\frac{m  |\Delta \vec x|^3  a^3 }{480   } \right]  \hspace{4 cm}\\
  \left.   +\frac{\ddddot a}{a}	\left[ -\frac{|\Delta \vec x|}{80 m }+\frac{|\Delta \vec x| \xi}{16 m }\right]  \right\} \ .
 \hspace{4 cm}\eea

On the other hand, the DeWitt-Schwinger calculation provides (\ref{result}). To compare it with the above result just expand it up to fourth adiabatic order (for the following identities we shall use the auxiliary parameter $T$ to denote the number of time-derivatives that are present). Use
\bea
g=1+\frac{1}{3}R_{\alpha\beta}y^{\alpha}y^{\beta}+\frac{1}{6}R_{\alpha\beta; \gamma}y^{\alpha}y^{\beta}y^{\gamma}+\left[\frac{1}{18}R_{\alpha\beta}R_{\gamma\delta}-\frac{1}{90}R_{\lambda\alpha\beta}^{\hspace{0.55 cm}k}R^{\lambda}_{\hspace{0.17cm}\gamma \delta k}+\frac{1}{20}R_{\alpha\beta;\gamma\delta} \right]y^{\alpha}y^{\beta}y^{\gamma}y^{\delta}+O(T^{-5}) \ , \nonumber\\
\eea
and relations  (\ref{F2})-(\ref{F1})   including the fourth adiabatic contributions \cite{brewin} (for simplicity we only show here, without loss of generality, the corresponding expressions for $\Delta t=0$)
\bea
y^0 & = & \frac{1}{2}H a^2 \Delta x^2+\frac{1}{144} a^4 \Delta x^4 H R +O(T^{-5}) \ ,\\
y^i & = & a\Delta x^i \left\{  1+\frac{1}{6} a^2 \Delta x^2 H^2+\frac{1}{120} a^4 \Delta x^4 H^2 \left[H^2+3\frac{\ddot a}{a} \right] 	\right\} +O(T^{-5})\ , \\
-2\sigma & = & \Delta x^2 a^2+\frac{1}{12 }a^4 \Delta x^4 H^2 + \frac{1}{360} a^6 \Delta x^6 H^2 \left[ H^2+3\frac{\ddot a}{a}\right] +O(T^{-5}) \ .
\eea
Using all these auxiliary expressions we can prove that the adiabatic scheme generates the same two-point function as the DeWitt-Schwinger one up to fourth order in the derivatives of the metric,
\bea \label{eqdsad4}
{^{(4)}}G_{DS}(x,x')={^{(4)}}G_{Ad}(x,x') \ .
\eea
We believe that  one might extend this identity up to any order by induction.
We note in passing that the result (\ref{eqdsad4}) implies the equivalence of the renormalized stress-energy tensor for 
$\xi \neq 1/6$ (see Sec. II D).  
\section{Conclusions and final comments}

The main motivation of this paper is to show the equivalence of the renormalized expectation values of the stress-energy tensor for spin-$1/2$ fields using both adiabatic and DeWitt-Schwinger  methods. This is a very natural question since the adiabatic renormalization scheme for Dirac fields has been introduced very recently in the literature. The  employed strategy  to achieve our goal has led us to show the equivalence for scalar fields as well, in a simpler way to that used in \cite{Birrell78, Anderson-Parker87}. Moreover, we were  naturally led to investigate the equivalence for the two-point function at coincidence  for both DeWitt-Schwinger and adiabatic series expansion at any  order. We have checked explicitly that the equality holds at sixth adiabatic order and we have argued that the equivalence must hold at an arbitrary order. This way,   the adiabatic regularization method will offer a very efficient computational tool to evaluate the higher order DeWitt coefficients in FLRW space-times for both scalar and Dirac fields. This may be relevant to capture nonperturbative aspects of the effective action in cosmological space-times, as those found in \cite{parker-toms85, parker-ravel, parker-vanzella}. Finally, we would like to remark that these results suggest that the equality ${^{(n)}}G_{Ad}(x, x)= {^{(n)}}G_{DS}(x, x)\ , \  n = 0, 2, 4, 6, ...$ (and the analogue for Dirac fields) could even hold
for separate points. This is actually supported by the fact that (\ref{equivalenceG4}), (\ref{equiSDSAd}), extended to separate points, 
coincide at  least  up to the fourth adiabatic order. 

\section*{Acknowledgments}
  
This work is supported  by the  Research Project of the Spanish MINECO   FIS2011-29813-C02-02,  
and the Consolider Program No. CPANPHY-1205388.
A. D. is supported by the Spanish Ministry of Education Ph.D. fellowship FPU13/04948. We would like to thank I. Agullo, A. Landete, F. Torrenti, and L. Parker for very useful discussions.

\appendix

\section{ADIABATIC EXPANSION FOR KLEIN-GORDON FIELDS}

In this section we show the generic expression for the $n$th contribution in the WKB adiabatic expansion given by (\ref{adiabatic}). Introducing the ansatz (\ref{kgansatz}) into the equation of motion for the modes, one finds the following equation \cite{parker-toms,birrell-davies},
\begin{eqnarray}
W_k^2=\omega^2+\sigma+\frac{3}{4}\frac{\dot{W_k}^2}{W_k^2}-\frac{1}{2}\frac{\ddot{W_k}}{W_k} \ , \label{A1}
\end{eqnarray}
where
\begin{eqnarray}
\sigma=\left(6\xi-\frac{3}{4}\right)\left(\frac{\dot{a}}{a}\right)^2+\left(6\xi-\frac{3}{2}\right)\frac{\ddot{a}}{a} \ .
\end{eqnarray}
Equation (\ref{A1}) can be solved {\it algebraically} by iteration for initial value $\omega^{(0)}\equiv\omega=\sqrt{(k/a)^2+m^2}$.  Performing the calculation up to $n$th adiabatic order it can be shown that
\begin{eqnarray}
\omega^{(n)} & = & \frac{1}{2\omega^3}\left \{  \omega^2\left[ (\omega^{(n/2)})^2 +2\sum_{i=2}^{n/2-1}\omega^{(i)} \omega^{(n-i)} \right]  +\sigma\left[  (\omega^{(n/2-1)})^2+2\sum_{i=0}^{n/2-2}\omega^{(i)} \omega^{(n-2-i)}   \right] \right.    \label{A}  \\
& & +\frac{3}{4}\left[  (\dot{\omega}^{(n/2-1)})^2+2\sum_{i=0}^{n/2-2}\dot{\omega}^{(i)} \dot{\omega}^{(n-2-i)}   \right] - \frac{1}{2}\left[ \ddot{\omega}^{(n/2-1)}\omega^{(n/2-1)}+ \sum_{i=0}^{n/2-2}\left(\ddot{\omega}^{(i)} \omega^{(n-2-i)} +\omega^{(i)} \ddot{\omega}^{(n-2-i)}  \right)  \right] \nonumber  \\
& & -\left[ 6 \sum_{i=0}^{n/4-1}(\omega^{(i)})^2 (\omega^{(n/2-i)})^2 +4 \sum_{k=0}^{n/2-1}(\omega^{(k)})^2 \sum_{i=2}^{n/2-k-1}\omega^{(i)}\omega^{(n-i-2k)}+4 \sum_{k=2}^{n/2-1}(\omega^{(k)})^2 \sum_{i=0}^{n/2-k-1}\omega^{(i)}\omega^{(n-i-2k)}\right.   \nonumber\\
& & \left. \left. +8\sum_{i=0}^{n/4-2}\omega^{(i)}\sum_{j=i+2}^{n/2-2-i}\omega^{(j)}\sum_{k=j+2}^{n-j-2i-2}\omega^{(k)}\omega^{(n-k-i-j)}+8\sum_{k=0}^{n/4-3/2}(\omega^{(k)})^2\sum_{i=k+2}^{n/2-k-1}\omega^{(i)}\omega^{(n-i-2k)} +(\omega^{(n/4)})^4 \right] \right\}  \ , \nonumber
\end{eqnarray}
with $\omega^{(s)}=0$ for $s<0$ or $s$ being a fractional number. With this formula we can recover $\omega^{(s)}=0$, for $s$ being an odd integer, and the corresponding expressions for orders $2$ and $4$ from \cite{parker-toms,birrell-davies}, 
\begin{eqnarray}
\omega^{(2)} & = & \frac{1}{2}\omega^{-1/2}\frac{d^2}{dt^2}\omega^{-1/2}+\frac{1}{2}\omega^{-1}\sigma	\ ,\\
\omega^{(4)} & = & \frac{1}{4}\omega^{(2)}\omega^{-3/2}\frac{d^2}{dt^2}\omega^{-1/2}-\frac{1}{2}\omega^{-1}(\omega^{(2)})^2 -\frac{1}{4}\omega^{-1/2}\frac{d^2}{dt^2}\left[\omega^{-3/2}\omega^{(2)}\right]	\ ,
\end{eqnarray}
as well. In general, (\ref{A}) allows us to obtain any $\omega^{(n)}$ in terms of lower order adiabatic terms and its derivatives.

\section{ADIABATIC EXPANSION FOR DIRAC FIELDS}

In this section we present the generic expressions for the $n$th contribution in the Dirac adiabatic expansion given by (\ref{fermion-ansatz})-(\ref{adifexp}). Introducing these expressions into the equation of motion for the modes, (\ref{3b}), one gets a set of coupled {\it algebraic} equations \cite{valencia}
\bea
(\omega-m)G & = & (\Omega-\omega)F+i \dot F-\frac{i m \dot \omega}{2\omega (\omega+m)}F+(\omega-m)F	\ ,\\
(\omega+m)F & = & (\Omega-\omega)G+i \dot G+\frac{i m \dot \omega}{2\omega (\omega-m)}G+(\omega+m)G		\ ,\\
2\omega & = &  (\omega+m)F F^*+(\omega-m) G G^*	\ ,
\eea
which can be solved {\it algebraically} by iteration for initial values $F^{(0)}=G^{(0)}=1$ and $\omega^{(0)}=\omega$. The general algorithm to compute the three fundamental objects [notice that $G(-m)$ satisfies the same equations as $F(m)$, so we take $G(-m)=F(m)$] is provided by 
\bea
\omega^{(n)} & = & - \frac{m}{\omega}	\left\{ \sum_{l=1}^{n-1}\omega^{(l)}F^{(n-l)}+i \dot F^{(n-1)}- \frac{i m \dot \omega}{2\omega(\omega+m)}  F^{(n-1)} \right\}  \label{B1} 	\\
& + &  \left(1-\frac{m}{\omega}\right) \left\{ -\frac{i}{2}\left[ \dot F^{(n-1)}+\dot G^{(n-1)}\right] -\frac{1}{2} \sum_{l=1}^{n-1}\omega^{(l)}\left[  F^{(n-l)}+ G^{(n-l)}  \right] + \frac{i m \dot \omega}{4\omega}\left[ \frac{F^{(n-1)}  }{(\omega+m)}- \frac{G^{(n-1)} }{(\omega-m)} \right]  \right\}	\ ,\nonumber	\\
Re\, F^{(n)}(m) & = & \frac{\delta_{n0} }{2}-\frac{1}{4\omega} \sum_{l=1}^{n-1}\left[ F^{(l)}F^{*(n-l)}(\omega+m)+G^{(l)}G^{*(n-l)}(\omega-m) \right] +\frac{1}{2\omega} Im\, \dot F^{(n-1)}(m)\nonumber\\
 &  & -\frac{1}{2\omega} \sum_{l=1}^{n} \omega^{(l)}Re\, F^{(n-l)} (m) -\frac{m \dot \omega}{4\omega^2 (m+\omega)}Im\, F^{(n-1)} (m) \ , \label{B2} \\
 Im\, F^{(n)}(m) & = & Im\, G^{(n)}(m) -\frac{1}{\omega-m}\left\{\sum_{l=1}^n \omega^{(l)} Im\, F^{(n-l)}(m)+Re\, \dot F^{(n-1)}(m)-\frac{m \dot \omega}{2\omega (\omega+m)} Re\, F^{(n-1)}(m) \right\} \ , \label{B3}
\eea 
with $F=Re\, F(m)+i Im\, F(m)$ and $G=Re\, G(m)+i Im\, G(m)$. Notice that there is an inherent ambiguity in the formalism reflected in the choice for $Im \, G(m)$, but it can be explicitly seen that it does not affect the observables such as  $\langle \bar \psi \psi \rangle$ or $\langle T_{\mu\nu} \rangle$ \cite{valencia}.  The simplest way to remove the ambiguities is to  assume $Im\,G^{(n)}(m)=-Im\, F^{(n)}(m)$. Detailed expressions for the first adiabatic contributions can be found in \cite{valencia, valencia2}.  In general, (\ref{B1})-(\ref{B3}) allow us to obtain any Dirac adiabatic contribution in terms of lower order adiabatic terms and its derivatives.

\section{$a_4$ COEFFICIENT. }
We give here the result for the $a_4$ DeWitt coefficient for a spatially flat  FLRW spacetime obtained with the adiabatic regularization method (\ref{A}):

 \bea a_4(x) & = & \frac{29 \dot a^8}{120a^8}-\frac{379\dot a^6 \ddot a}{210 a^7}+\frac{899 \dot a^4 \ddot a^2}{280 a^6}+\frac{83\dot a^2 \ddot a^3}{35a^5}-\frac{13 \ddot a^4}{21a^4}+\frac{47 \dot a^5 \dddot a}{70 a^6}+\frac{2\dot a^3 \ddot a\dddot a}{3 a^5}-\frac{103 \dot a \ddot a^2\dddot a}{28 a^4}-\frac{647\dot a^2 \dddot a^2}{840 a^4}\nonumber\\ & &+\frac{103 \ddot a^2 \ddddot a}{210 a^3}-\frac{2\dot a^4 \ddddot a}{21a^5}-\frac{93 \dot a^2\ddot a \ddddot a}{70 a^4}+\frac{34 \ddot a^2 \ddddot a}{105 a^3}+\frac{199\dot a \dddot a\ddddot a}{420 a^3}+\frac{11 \ddddot a^2}{504 a^2}-\frac{13\dot a^3 a^{(5)} }{210 a^4}+\frac{41\dot a \ddot a  a^{(5)} }{140 a^3}\nonumber\\  && +\frac{29 \dddot a  a^{(5)}}{1260 a^2}+\frac{3\dot a^2  a^{(6)}}{70 a^3}+\frac{13 \ddot a  a^{(6)}}{1260 a^2}-\frac{\dot a  a^{(7)}}{126a^2}-\frac{ a^{(8)}}{630a}-\frac{7\xi \dot a^8}{5 a^8}-\frac{39\xi^2 \dot a^8}{5 a^8}+\frac{36\xi^3 \dot a^8}{a^8}+\frac{54\xi^4 \dot a^8}{a^8}\nonumber \\  & & +\frac{383\dot a^6 \ddot a}{20 a^7}-\frac{15\xi^2 \dot a^6 \ddot a}{a^7}-\frac{234\xi3 \dot a^6 \ddot a}{a^7}+\frac{216\xi^4\dot a^6 \ddot a}{ a^7}- \frac{8123\xi \dot a^4 \ddot a^2}{140 a^6}+\frac{2859\xi^2\dot a^4 \ddot a^2}{10 a^6} \nonumber \\   & &-\frac{432\xi^3 \dot a^4 \ddot a^2}{ a^6}+\frac{324\xi^4 \dot a^4 \ddot a^2}{ a^6}-\frac{254\xi \dot a^2 \ddot a^3}{15 a^5}+\frac{264 \xi^2 \dot a^2 \ddot a^3}{5 a^5}-\frac{180\xi^3 \dot a^2 \ddot a^3}{a^5}+\frac{216 \xi^4\dot a^2 \ddot a^3}{a^5}+\frac{523\xi \ddot a^4}{105 a^4}\nonumber\\   &&-\frac{81\xi^2 \ddot a^4}{10 a^4}-\frac{18\xi^3 \ddot a^4}{a^4} +\frac{54\xi^4 \ddot a^4}{a^4}-\frac{211\xi \dot a^5 \dddot a}{20 a^6}+\frac{201\xi^2 \dot a^5 \dddot a}{5 a^6}-\frac{18\xi^3 \dot a^5 \dddot a}{ a^6}+\frac{53\xi \dot a^3 \ddot a\dddot a}{7 a^5}-\frac{69\xi^2\dot a^3 \ddot a\dddot a}{a^5}\nonumber \\   && +\frac{72\xi^3\dot a^3 \ddot a\dddot a}{ a^5}+\frac{439\xi \dot a \ddot a^2\dddot a}{14 a^5}-\frac{84\xi^2\dot a \ddot a^2\dddot a}{ a^4}+\frac{90\xi^3\dot a \ddot a^2\dddot a}{ a^4}+\frac{6\xi \dot a^2 \dddot a^2}{a^4}-\frac{147\xi^2\dot a^2 \dddot a^2}{10 a^4} \nonumber\\   && +\frac{18\xi^3\dot a^2 \dddot a^2}{ a^4}-\frac{51\xi^3\ddot a \dddot a^2}{20 a^3}-\frac{3\xi^2\ddot a \dddot a^2}{ a^3}+\frac{18\xi^3\ddot a \dddot a^2}{ a^3}+\frac{11\xi\dot a^4 \ddddot a}{4 a^5}-\frac{15\xi^2\dot a^4 \ddddot a}{a^5}+\frac{18\xi^3\dot a^4 \ddddot a}{ a^5}\nonumber\\   &&+\frac{157\xi\dot a^2 \ddot a \ddddot a}{14 a^4}-\frac{153\xi^2\dot a^2 \ddot a \ddddot a}{5 a^4} +\frac{36\xi^3\dot a^2 \ddot a \ddddot a}{ a^4}-\frac{19\xi \ddot a^2 \ddddot a}{14 a^3}-\frac{24\xi^2 \ddot a^2 \ddddot a}{5 a^3}+\frac{18\xi^3 \ddot a^2 \ddddot a}{ a^3}  \nonumber\\   &&- \frac{237\xi \dot a \dddot a \ddddot a}{70 a^3}+ \frac{27\xi^2 \dot a \dddot a \ddddot a}{5 a^3}-\frac{39\xi \ddddot a^2}{140 a^2} + \frac{9\xi^2 \ddddot a^2}{10 a^2}+\frac{3\xi \dot a^3 a^{(5)}  }{10 a^4} +\frac{41 \dot a \ddot a a^{(5)}  }{140 a^3}-\frac{15\xi \dot a\ddot a a^{(5)}  }{7a^3} \nonumber\\  && +\frac{18\xi^2 \dot a\ddot a a^{(5)}  }{5 a^3}-\frac{12\xi \dddot a a^{(5)}  }{35 a^2}+\frac{6\xi^2 \dddot a a^{(5)}  }{5 a^2}+\frac{3 \dot a^2 a^{(6)}  }{70 a^3}-\frac{23\xi \dot a^2 a^{(6)}}{70 a^3}+\frac{3\xi^2 \dot a^2 a^{(6)}}{5 a^3}-\frac{6\xi \ddot a a^{(6)}}{35 a^2}+\frac{3\xi^2 \ddot a a^{(6)}}{5 a^2}\nonumber\\   &&+\frac{\xi \dot a a^{(7)}}{28 a^2}+\frac{\xi  a^{(8)}}{140 a^2} \ . \eea

\end{document}